\documentclass[12pt,preprint]{revtex4-1}

\usepackage{graphicx}
\usepackage{epstopdf}
\newcommand{\mJ}{\mathcal{J}}
\newcommand{\PJ}{P_\mathcal{J}}
\newcommand{\XJ}{X_\mathcal{J}}

\begin{document}

\title{A numerical study of the overlap probability distribution and
  its sample-to-sample fluctuations in a mean-field model}

\author{Giorgio Parisi} \affiliation{Dipartimento di Fisica,
  INFN -- Sezione di Roma 1 and CNR -- IPCF, UOS di Roma,\\
  Universit\`{a} La Sapienza, P.le A. Moro 5, 00185 Roma, Italy}

\author{Federico Ricci-Tersenghi} \affiliation{Dipartimento di Fisica,
  INFN -- Sezione di Roma 1 and CNR -- IPCF, UOS di Roma,\\
  Universit\`{a} La Sapienza, P.le A. Moro 5, 00185 Roma, Italy}

\date{\today}

\begin{abstract}
In this paper we study the fluctuations of the probability distributions of the overlap in mean field spin glasses in the presence of a magnetic field on the De Almeida-Thouless line.
We find that there is a large tail in the left part of the distribution that is dominated by the contributions of rare samples. Different techniques are used to examine the data and to stress on different aspects of the contribution of rare samples.

\end{abstract}

\maketitle

\section{Introduction}

Spin glass models show amazing physical properties.  Let us
considering for simplicity mean field spin glasses, like the
Sherrington-Kirkpatrick model \cite{SK}, whose solution is given by
the hierarchical replica symmetry breaking (RSB) Ansatz \cite{mpv,parisibook2,mpv,G,TALE}.
The model is defined by the following Hamiltonian
\begin{equation}
\mathcal{H}[\vec\sigma] = -\sum_{i,j} J_{ij} \sigma_i \sigma_j\;,
\label{ham}
\end{equation}
where $\sigma_i=\pm1$ are $N$ Ising spins and the $J_{ij}$ are
quenched random couplings with zero mean and variance $1/N$.

For each sample $\mJ$, that is for a choice of the quenched random
couplings, one can compute the probability distribution function of
the overlap, $q=\sum_{i=1}^N \sigma_i \tau_i / N$, between two
replicas $\vec\sigma$ and $\vec\tau$ subject to the same Hamiltonian
(\ref{ham}): we call $\PJ(q)$ such a probability distribution.

In the SK model the order
parameter in the thermodynamical limit is given by a function
$q(x):[0,1]\to[0,1]$, related to the probability distribution $P(q)$ of
finding two replicas at an overlap $q$, where $P(q)$ is defined as
\[
P(q) = \overline{\PJ(q)}\;,
\]
and the overline represents the average over the samples $\mJ$'s.

The overlap distribution $\PJ(q)$ strongly fluctuates from sample to
sample. In the low temperature spin glass phase ($T< T_c$), these
distributions are not self-averaging, that is the typical $\PJ(q)$ is
very different from the disorder averaged distribution $P(q)$, even in
thermodynamical limit.  The size of these fluctuations in the SK model
can be quantified by using  the Ghirlanda-Guerra relations \cite{RUELLETREE,mpv,GG,AI,SOL}; the simplest identity  is
\[
\overline{\PJ(q) \PJ(q')}-\overline{\PJ(q)}\;\overline{\PJ(q')} =
\frac13 \Big[\delta(q-q') - P(q) \Big] P(q')\;,
\]
and the r.h.s.\ is non null as soon as the $P(q)$ is not a delta
function, i.e.\ when replica symmetry is broken.

These large sample-to-sample fluctuations play a very relevant role in
numerical simulations, since they require a huge number of samples to
obtain reliable measurements in the low temperature phase of spin
glass models, and they may produce finite size effects that vanish
very slowly, increasing system size.

In the present paper we study overlap distributions in a mean-field
spin glass model, defined on a Bethe lattice of fixed degree, in the
presence of an external field. We focus on the data measured at the
critical temperature $T_c$, such that the mean overlap distribution in
the thermodynamical limit is a delta function,
$P(q)=\delta(q-q_0)$. This choice has two main advantages:
\begin{itemize}
\item we know analytically the value of $T_c$ and $q_0$, by solving
  the model with the cavity method, and this allows us to better study
  deviation from the thermodynamical limit (i.e., finite size
  effects);
\item the system is critical and so it shows very large sample to
  sample fluctuations.
\end{itemize}
For any temperature different from $T_c$ one of the two above
statement would be false, thus making our study less interesting.
Moreover the presence of the external field breaks the global spin
inversion symmetry and implies that overlaps are non-negative in the
thermodynamical limit: however it is well known that a large tail in
the negative overlap region is present in systems of finite size and
its origin needs to be clarified.

\section{The model and the numerical simulations}

We study an Ising spin glass model defined on a Bethe lattice of fixed
connectivity $c=4$ (i.e., a random regular graph of fixed degree
$c=4$). The Hamiltonian is
\begin{equation}
\mathcal{H} = -\sum_{<ij>} J_{ij} \sigma_i \sigma_j - H \sum_i \sigma_i\;,
\label{hamBethe}
\end{equation}
where $\sigma_i=\pm1$ are $N$ Ising spins, the couplings $J_{ij}=\pm1$
(with equal probability) are quenched random variables and the sum
runs over all pairs of neighboring vertices in the graph. We use a
constant external field $H > 0$. For not very small connectivity Ising spin glasses on a Bethe lattice share many properties with the Sherrrington-Kirkpatrick model \cite{VB,BETHE}: in the limit $c\to \infty$ we recover the Sherrington Kirkpatrick model, and the $1/c$ corrections are well under control \cite{ParisiTria}.

We construct the random regular graph  in the following way: we
attach $c$ legs to each vertex and then we recursively join a pair of
legs, forming a link, until no legs are left or a dead end is reached
(this may happen because we avoid self-linking of a vertex and
double-linking between the same pair of vertices); if a dead end is
reached, the whole construction is started from scratch.

\begin{figure}[htb]
\begin{center}              
\includegraphics[width=0.7\textwidth]{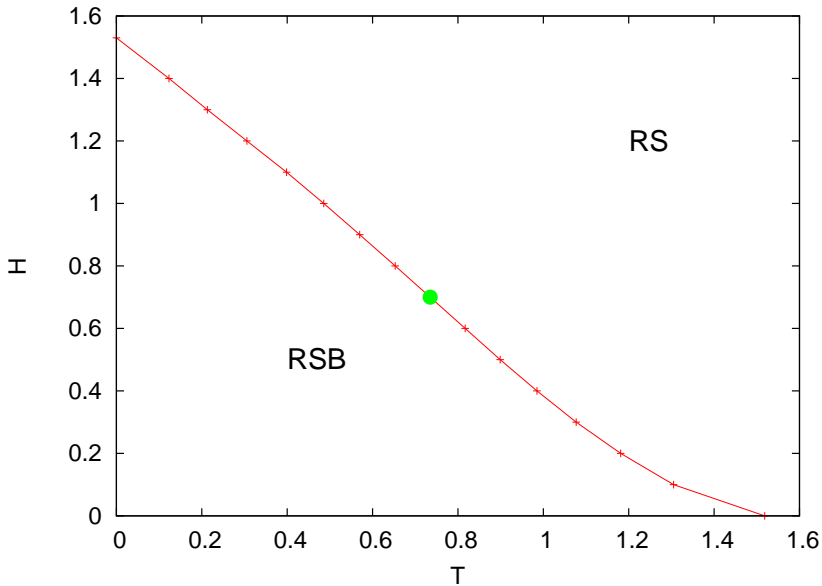}
\caption{Phase diagram in the temperature--field plane for the
  $J=\pm1$ spin glass model defined on a Bethe lattice of fixed degree
  $c=4$. In this work we report data collected at the critical point
  marked by a big dot.}
\label{phaseDiagram}
\end{center} 
\end{figure}

Similarly to the SK model, the model (\ref{hamBethe}) has a continuous spin
glass phase transition at a critical temperature $T_c$ which depends
both on the value of $c$ and $H$. At variance with the SK model, the
critical line in the $(T,H)$ plane does not diverge when $T\to 0$, but
rather reaches a finite value $H_c$ (see Fig.~\ref{phaseDiagram}).
This is due to the finite number of neighbors each spin has on a
random graph of finite mean degree (while this number is divergent
with the system size in the SK model). In this sense the present model
is closer to finite dimensional models than the SK model is.

The replica symmetric (RS) phase of model (\ref{hamBethe}) can be solved
analytically by the cavity method \cite{BETHE}. In particular one can find the
boundary of the RS phase, beyond which the model solution
spontaneously breaks the replica symmetry \cite{PaPaRa,TaRiKa}. In Fig.~\ref{phaseDiagram}
we show such a critical line in the $(T,H)$ plane for the model with
fixed degree $c=4$.  The high temperature and/or high field region is
replica symmetric, while a breaking of the replica symmetry is
required in the low temperature and low field region. We have checked
that the phase boundary behaves like $H_c(T) \propto (T-T_c)^{3/2}$
close to zero-field critical point $T_c$, and the exponent is the same
one found in the SK model.

We have carried our Monte Carlo simulations at the point marked with a
big dot in Fig.~\ref{phaseDiagram}, that is $H=0.7$ and $T=0.73536$.
The uncertainty on the critical temperature for $H=0.7$ is $10^{-5}$.
At that point the value of the thermodynamical overlap is
$q_0=0.67658(1)$.  Please notice that we have chosen a rather large
value of the external field, which is roughly half of the largest
critical field value $H_c(T=0) \simeq 1.53$, in order to avoid
crossover effects that could be due to the vicinity of the zero-field
critical point.

\begin{figure}[htb]
\begin{center}              
\includegraphics[width=0.7\textwidth]{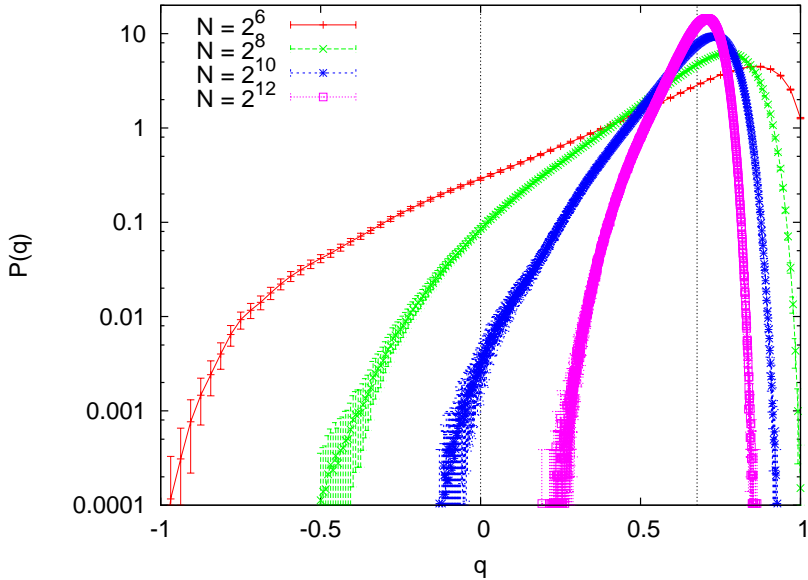}
\caption{Disorder averaged overlap probability distributions $P(q)$
  show an exponential tail for $q < q_0$.}
\label{Pq}
\end{center}
\end{figure}

Monte Carlo simulations have been performed by using the Metropolis
algorithm and the parallel tempering method: we used 20 temperatures
equally spaced between $T_{max}=2.0$ and $T_c=0.73536$, and we
attempted the swap of configurations at nearest temperatures every 30
Monte Carlo sweeps (MCS). Each sample (of any size) has been
thermalized for $2^{24}$ MCS and then 1024 measurements have been
taken during other $2^{26}$ MCS: so there are $2^{16}$ MCS between two
successive measurements and we have checked this number to be larger
than the autocorrelation time. We study systems of sizes ranging from
$N=2^6$ to $N=2^{14}$, with a number of samples ranging from 5120 for
$N=2^6$ to 1280 for $N=2^{14}$. We are going to present only the data
for sizes $N \le 2^{12}$ for which we have simulated at least 2560
samples; indeed the data for $N=2^{13}$ and $N=2^{14}$ are more noisy (due to the limited number of samples),
 moreover we fear that some samples may not be perfectly
thermalized even after $2^{26}$ MCS. By restricting to $N \le 2^{12}$
we are fully confident about the numerical data.

\section{Results}

We start by showing in Fig.~\ref{Pq} the disorder averaged $P(q)$
for different sizes. The exponential tail on the left side is evident
from the plot (which is on a logarithmic scale): this tail goes far in
the negative overlap region for small sizes. In the following we are
going to show that this exponential tail is not a feature of typical
samples, but it is completely due to very rare and atypical samples.

\begin{figure}[htb]
\begin{center}
\includegraphics[width=0.49\textwidth]{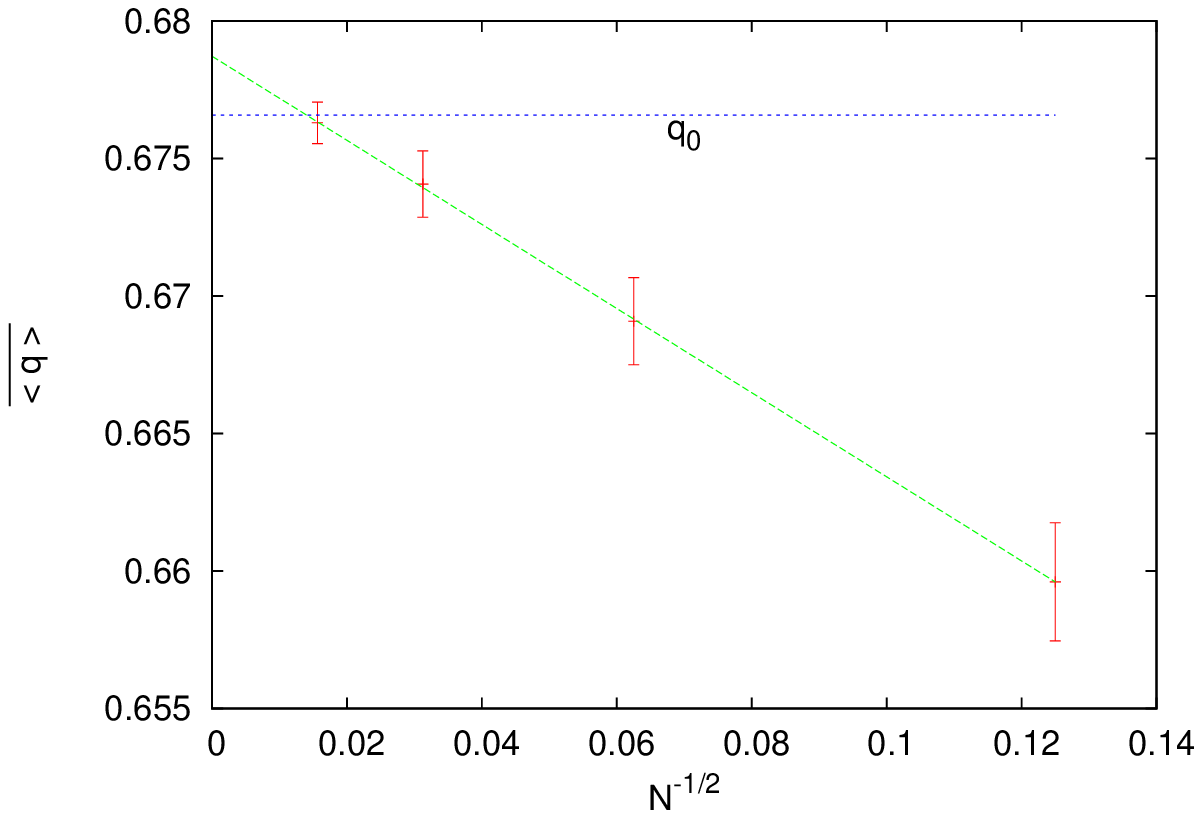}
\includegraphics[width=0.49\textwidth]{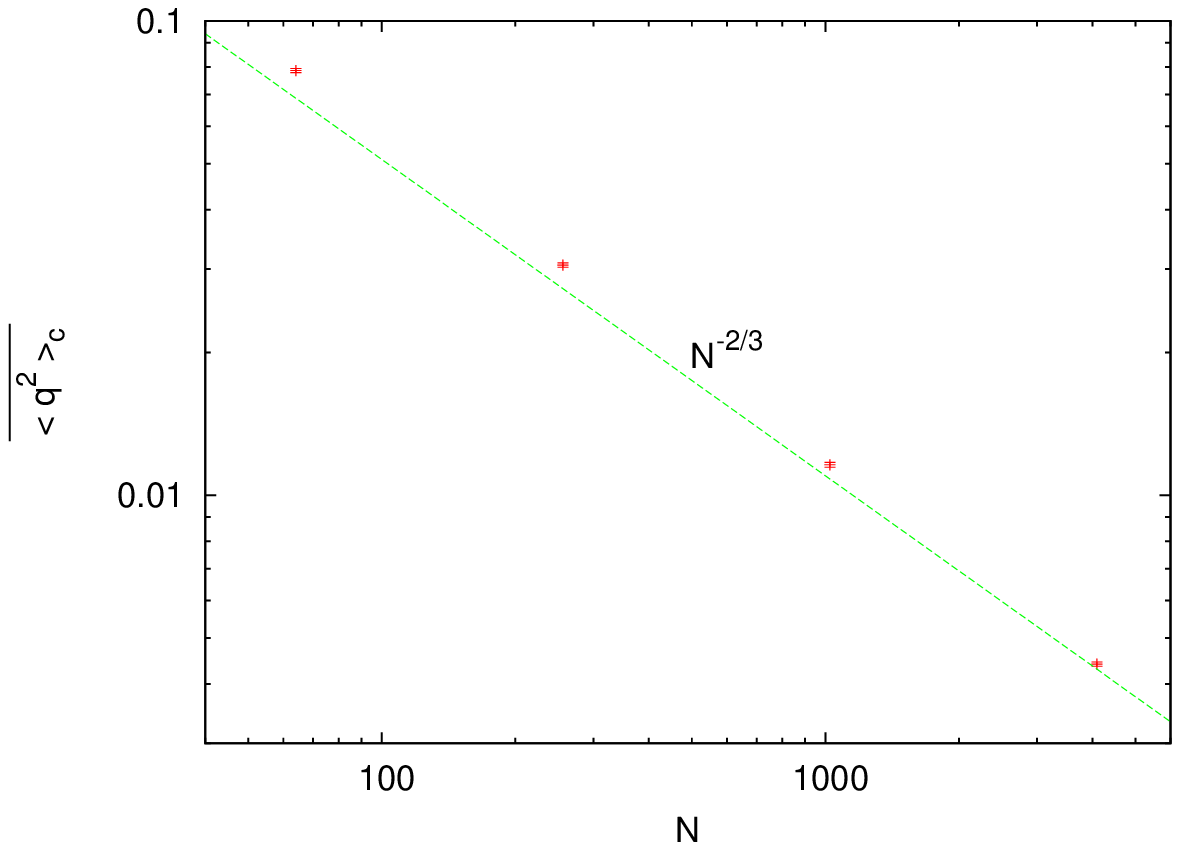}
\caption{Mean (left) and variance (right) of $P(q)$. A naive analysis
  would predict an asymptotic value for $\langle q \rangle$ larger
  than $q_0$ and finite size corrections decaying faster than the
  $N^{-1/3}$ expected behavior. The variance decays roughly as the
  expected $N^{-2/3}$ law.}
\label{mom}
\end{center} 
\end{figure}

The vertical line at $q=q_0$ in Fig.~\ref{Pq} marks the location of
the delta peak in the thermodynamical limit.  By looking at the mean
and the variance of $P(q)$ we have checked how finite size effects
decay to zero.  We see in Fig.~\ref{mom} that while $\overline{\langle
  q^2 \rangle}_c$ decays in a way compatible with the expected
behavior $N^{-2/3}$ (the discrepancy can be well explained in terms
of small scaling corrections), the mean overlap $\overline{\langle q
  \rangle}$ shows finite size corrections proportional to $N^{-1/2}$
(instead of the expected $N^{-1/3}$) and seems to extrapolate to a
thermodynamical limit different from $q_0$.  This means that a naive
extrapolation to the thermodynamical limit would produce a wrong
estimate of $q_0$.  The most probable explanation is that finite size
corrections of order $N^{-1/2}$ have a much larger coefficient than
those of order $N^{-1/3}$ and then much larger sizes are needed to
observe the asymptotic behavior.

\begin{figure}[htb]
\begin{center}              
\includegraphics[width=0.8\textwidth]{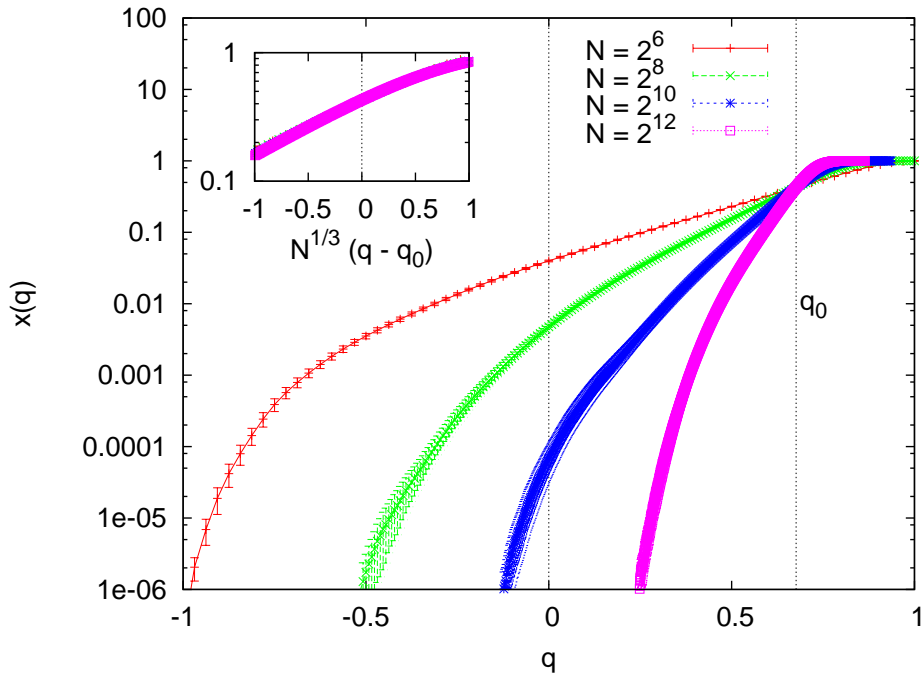}
\caption{The integrated probability distribution $x(q)$ averaged over
  the disorder. The value of $q_0$ can be well determined by the
  crossing point of data set in the main panel. The inset show the
  scaling function $x(N^{1/3}(q-q_0))$.}
\label{xq}
\end{center} 
\end{figure}

A much better way to estimate $q_0$ from the disorder averaged data
seems to be the analysis of the overlap integrated probability
function
\[
x(q) \equiv \int_{-1}^q P(q') dq'\;.
\]
This variable has been used for studying the behavior of three dimensional systems at zero magnetic field \cite{MaPaRu,Janus5}. The results are shown in Fig.~\ref{xq}.  In the main panel we see again
the exponential tail on the left side, but the crossing point of the
functions $x(q)$ estimates with a high accuracy the right value for
$q_0$.  In the present case all the crossing points for the sizes
shown are within a distance less than $10^{-3}$ from the
thermodynamical value and converge to it according to the $N^{-1/3}$
law. In the inset of Fig.~\ref{xq} we show that $x(q)$ data
perfectly collapse when plotted as a function of the scaling variable
$N^{1/3}(q-q_0)$.

Let us now turn to the study of sample-to-sample fluctuations.  We
want to convince the reader that the exponential tail is not a feature
of typical samples: actually not even a feature of the vast majority
of samples, that show roughly Gaussian (or even steeper) tails in
their $\PJ(q)$.  The exponential tail is produced by the integration
of the secondary peak that atypical samples have at an overlap value
much smaller than $q_0$.

\begin{figure}[htb]
\begin{center}              
\includegraphics[width=0.8\textwidth]{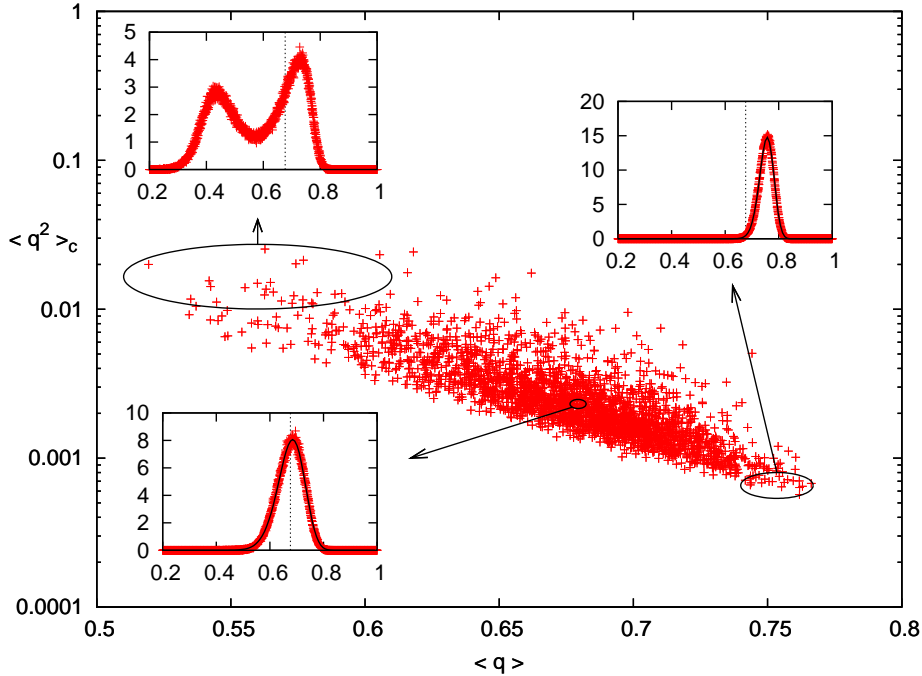}
\caption{Mean and variance of the 2560 samples of size
  $N=2^{12}$. Insets show the overlap probability distribution
  averaged over a small fraction, $1/128$, of samples (those in the
  corresponding circle). Solid curves in the insets are Gaussian fits
  to the data (see text for details).}
\label{q1q2Pq}
\end{center} 
\end{figure}

Extracting typical and atypical shapes of the $\PJ(q)$ from thousands
of samples is not a straightforward job. We follow the simplest
procedure based on the analysis of the first two moments.  In the main
panel of Fig.~\ref{q1q2Pq} we show the mean and the variance of the
2560 samples of size $2^{12}$.  The three insets in Fig.~\ref{q1q2Pq}
show the averages over 20 $\PJ(q)$ chosen from typical samples (lower
inset) or from atypical samples, either much broader or much narrower
than typical (upper insets). In every inset we also draw a dashed
vertical line to mark the location of $q_0$.

We notice that there exist a large difference between typical an
atypical samples, both quantitatively and qualitatively (especially
for the atypical samples showing a double peak structure). However the
very different shapes can be roughly accounted for by considering an
effective external field different from the one ($H=0.7$ in the
present case) appearing in the Hamiltonian: in the atypical samples
shown in the upper right inset this effective field is larger than $H$
and thus the overlap distribution is narrower and centered on a value
greater than $q_0$, while the atypical samples shown in the upper left
inset look like if they were below the critical line, i.e. with a
field smaller than $H$.

Since samples with different effective fields will have different
critical temperatures, it is possible that the main source of sample-to-sample
fluctuations can be well described by a random temperature (or field)
term in the effective Hamiltonian as in the case of ferromagnets in random magnetic field \cite{Sou,PaSou,PaPiSou}.

It is also worth noticing that the tails of the distributions shown in
the insets of Fig.~\ref{q1q2Pq} are Gaussian or even steeper, as  expected \cite{FraPaVi}. Indeed,
the interpolating curves superimposed to the bimodal distributions
(lower left and upper right insets) have been obtained by assuming
$q=\tanh(h)$ with a Gaussian distributed local field $h$. The
non-linear transformation is necessary (and sufficient) to take into
account the small skewness of the distributions.

\begin{figure}[htb]
\begin{center}              
\includegraphics[width=0.49\textwidth]{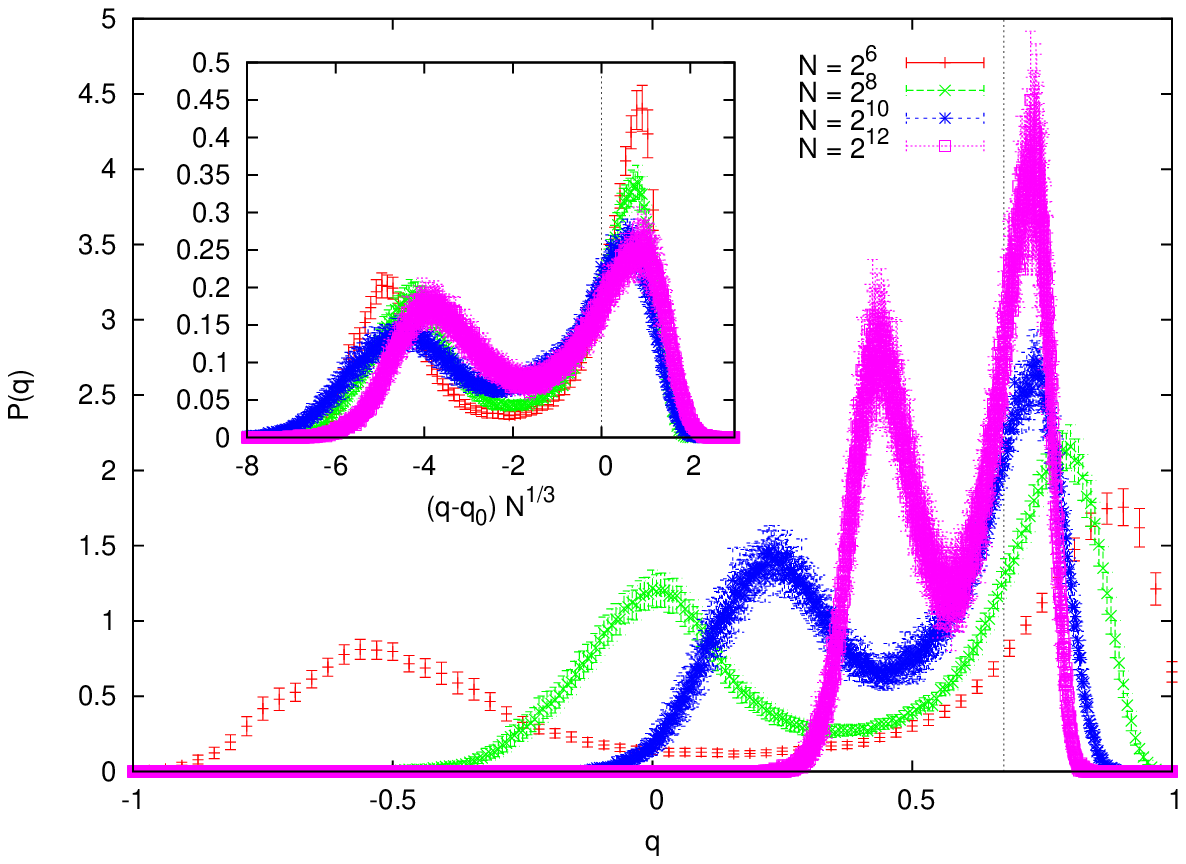}
\includegraphics[width=0.49\textwidth]{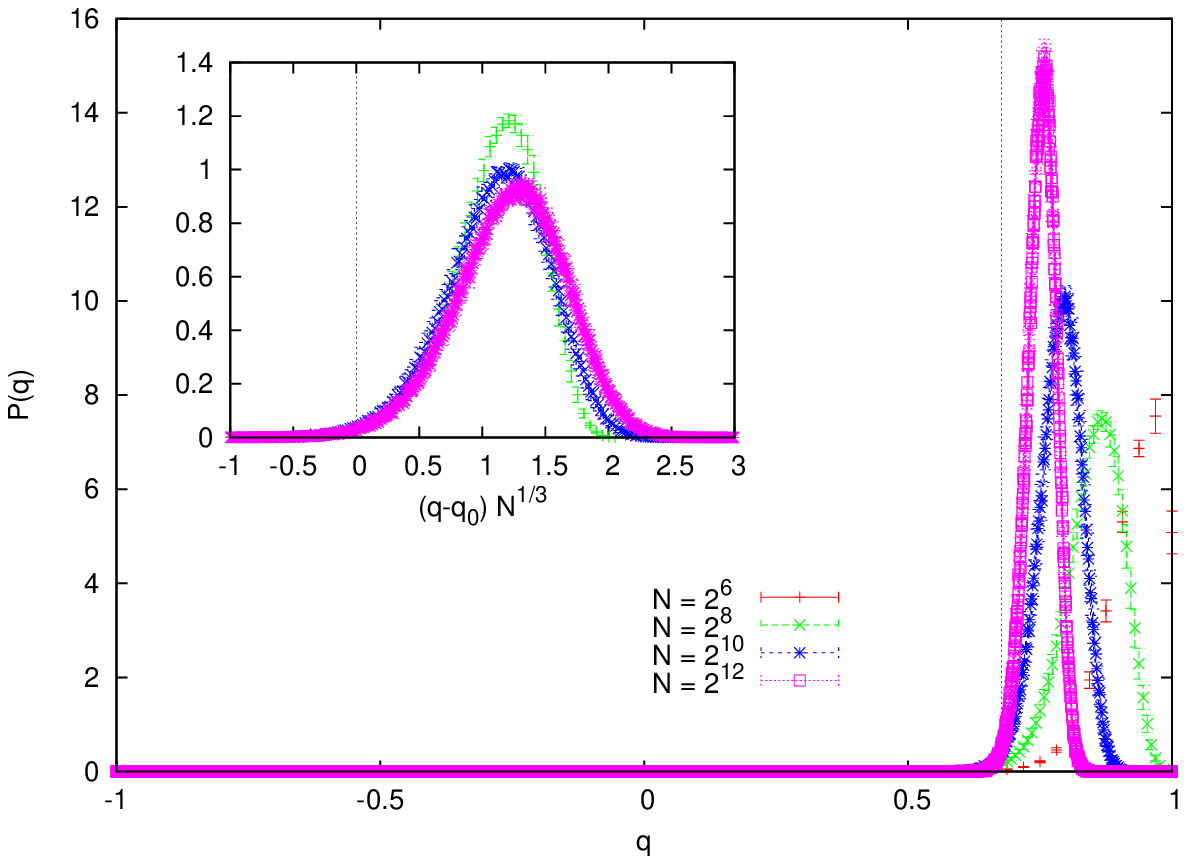}
\caption{Most atypical distributions, averaged over a fraction $1/128$
  of samples: those with the largest (left) and smallest (right)
  variance. By varying the system size they roughly preserve the shape
  and get shrunk according to the scaling $q-q_0 \sim N^{-1/3}$ (see
  insets).}
\label{atyp}
\end{center} 
\end{figure}

In Fig.~\ref{q1q2Pq} we have presented data only for size $N=2^{12}$,
but a natural question is how sample-to-sample fluctuations vary with
the system size. We have found that by increasing the system size the
distribution of the moments shrinks towards the thermodynamical limits
($\langle q \rangle = q_0$ and $\langle q^2 \rangle_c=0$) with the
expected $N^{-1/3}$ scaling behavior. However it is not true that all
samples become typical in the thermodynamical limit. In other words
the fraction of atypical samples (e.g.\ those with a bimodal
distribution) remains roughly constant.  In Fig.~\ref{atyp} we show
the average $P(q)$ computed on a small fraction (1/128) of samples,
those most atypical, i.e. those corresponding to upper insets in
Fig.~\ref{q1q2Pq}.  We notice that, by varying the size, the shape is
more or less preserved and the main effect is an overall shrink of the
distribution.  The insets in Fig.~\ref{atyp} show that this shrinking
is consistent with the scaling law $q-q_0 \sim O(N^{-1/3})$ that holds
at criticality.
\begin{figure}[htb]
\begin{center}              
\includegraphics[width=0.49\textwidth]{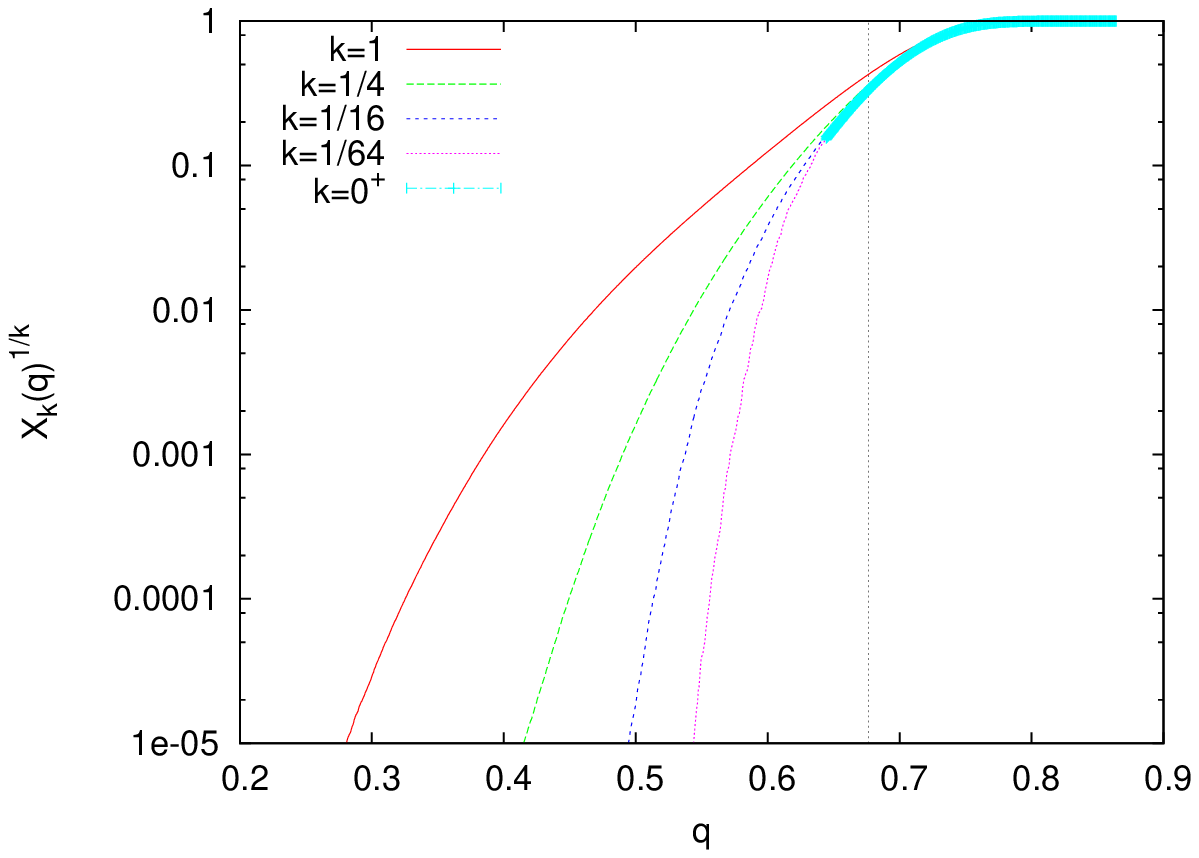}
\includegraphics[width=0.50\textwidth]{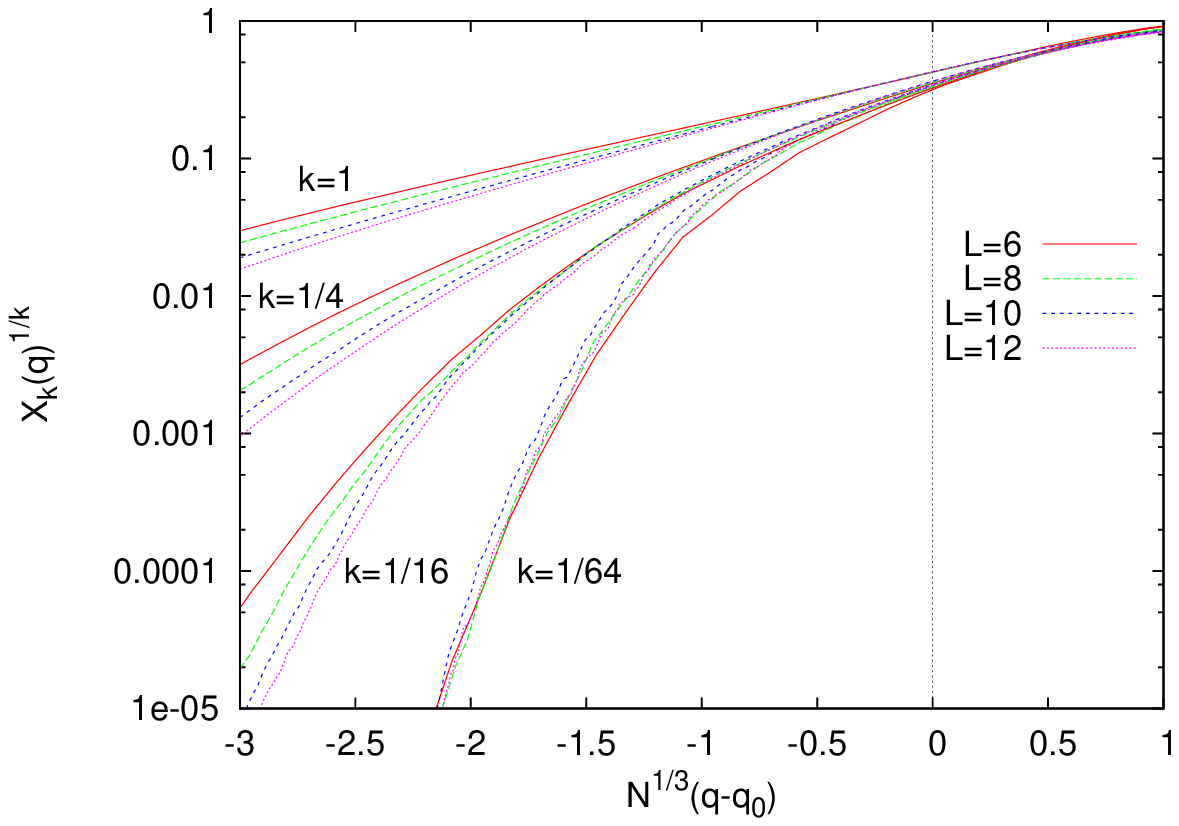}
\caption{Left: Small $k$ moments of $\XJ(q)$ measured in systems of
  size $N=2^{12}$ show a decay faster than exponential for $q \ll
  q_0$.  Right: Systems of different sizes show the same behavior,
  once the overlap is rescaled accordingly.}
\label{momX}
\end{center} 
\end{figure}

Given that neither typical nor atypical distributions have an
exponential tail, the only possible explanation is that such a tail is
generated by the secondary peak of broader distributions when
averaging over the samples.  We are going to provide quantitative
evidence for this by looking at the integrated probabilities
\[
\XJ(q) \equiv \int_{-1}^q \PJ(q') dq'\;.
\]
Let us define the moments of the random variable $\XJ$ as
\[
X_k(q) \equiv \overline{\XJ(q)^k}\;.
\]
Remind that $X_1(q)=x(q)$ is plotted in Fig.~\ref{xq} and shows an
exponential tail.  However in the region $q \ll q_0$ the average
$X_1(q)$ is dominated by rare samples, while the vast majority of
samples has a very small value $\XJ(q) \ll X_1(q)$ and do not
contribute to $X_1(q)$.

\begin{figure}[htb]
\begin{center}              
\includegraphics[width=0.8\textwidth]{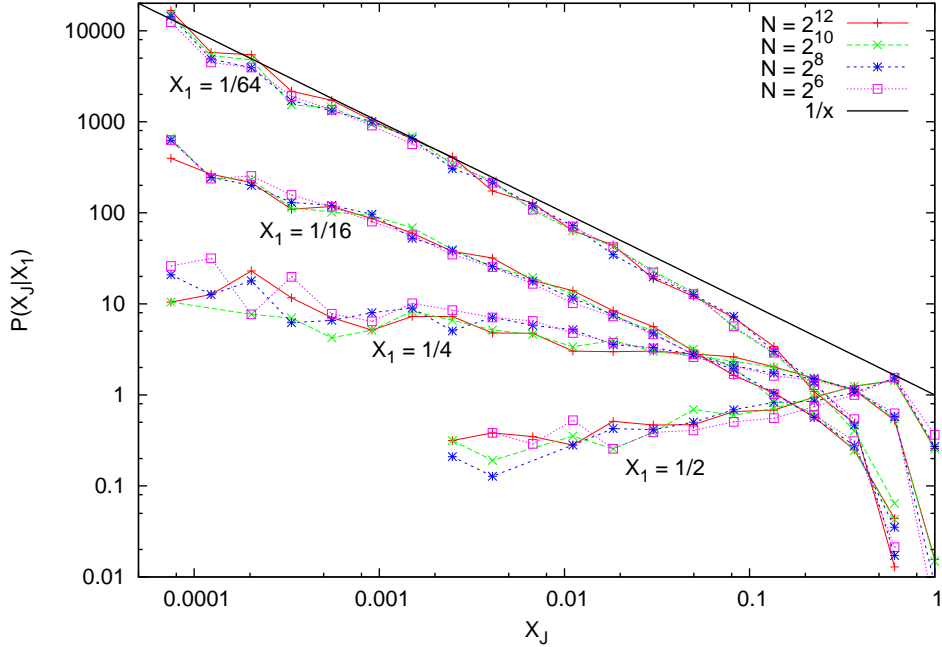}
\caption{Histograms of $\XJ$ at fixed values of $X_1$. The data for
  $X_1=1/64$ has been multiplied by 10 in order to improve
  readability.}
\label{histoLogX}
\end{center} 
\end{figure}

In order to extract the behavior of typical samples one should average
the random variable $\log(\XJ(q))$. However for $q \ll q_0$ there are
samples with $\XJ(q)=0$ and a straightforward computation of
$\overline{\log[\XJ(q)]}$ is not possible.  However, by noticing that
\[
\overline{\log[\XJ(q)]} = \lim_{k\to 0} \log\left[ X_k(q)^{1/k}
\right]\;,
\]
it is possible to observe the behavior of typical samples by choosing
$0 < k \ll 1$.  In Fig.~\ref{momX} (left) we plot $X_k(q)$ for
$k=1,1/4,1/16,1/64$ and we clearly see how the exponential tail for
$k=1$ becomes a Gaussian (or even steeper) decay for $k \ll 1$.
Moreover this behavior is very well conserved by varying the system
size: in Fig.~\ref{momX} (right) we plot the same averages, $X_1$ to
$X_{1/64}$, as a function of the scaling variables $N^{1/3}(q-q_0)$
and we see that the data collapse (which is very good for $q \simeq
q_0$) remains a reasonable approximation in the entire $q$ range.
This observation suggests that the entire distribution of the random
variable $\XJ(q)$ mainly depends on the scaling variable
$N^{1/3}(q-q_0)$, or equivalently on the first moment $X_1$.  This can
be checked in Fig.~\ref{histoLogX} where we plot the distribution of
$\XJ$ at some fixed value of $X_1$ for several system sizes.  Please
note that the data for $X_1=1/64$ have been multiplied by a factor 10
in order to avoid overlaps with other data set and improve
readability.

By commenting Fig.~\ref{histoLogX} we can finally draw the main
conclusions of this analysis.  First of all, the good data collapse for the probability
distribution of $\XJ$ at fixed value of $X_1$ is a strong indication
that we have measured large enough systems in the asymptotic scaling
regime. Moreover we see that for $X_1=1/2 > x(q_0) \simeq 0.429$ the
distribution of $\XJ$ has a maximum close to $X_1$, that is the mean
value is representative of typical samples behavior. On the contrary,
for $X_1 < x(q_0)$, the distributions of $\XJ$ have their maxima at
$\XJ \simeq 0$ and the mean value is not representative of typical
values. In particular we observe that for very small values of $X_1$
the distribution of $\XJ$ develops a power law divergence $1/\XJ$
for $\XJ \to 0$.
\section{Conclusions}

In this paper we have seen that on the De Almeida-Thouless line the left tail of the distribution of $P(q)$ is dominated by rare samples.
The presence of a left tail in the probability is quite an annoying phenomenon that is present also in the Sherrington-Kirpatrick model \cite{BiCo} and in finite dimensional models \cite{LePaRiRu}, both at the phase transition point and below the transition. This tail is particularly bothering at not too large magnetic field, because it extends in the region of negative $q$. We think that understanding the origin of this tail may be useful in future analysis of the finite dimensional simulations. 

It would be very useful to derive  the results on this paper in an analytic way extending  the techniques of  \cite{FraPaVi}. Indeed in that paper the computation of the tail was done for the typical samples and we have to modify it in order to compute the tail of the average over the samples. We believe that this is a feasible task.


\begin{thebibliography}{99}

\bibitem{SK} D. Sherrington and S. Kirkpatrick, Phys. Rev. Lett.  {\bf 35},  1792 (1975).

\bibitem{CINQUE}  E. Marinari, G. Parisi, F. Ricci-Tersenghi, J. Ruiz-Lorenzo, F. Zuliani
     J. Stat. Phys. {\bf 98},  973 (2000).

\bibitem{mpv} M. M\'ezard, G. Parisi and M.A. Virasoro, {\sl Spin glass theory and beyond}, World Scientific (Singapore 
1987).

\bibitem{parisibook2} G.Parisi, {\sl Field Theory, Disorder and Simulations}, World Scientific, (Singapore 1992).

\bibitem{G} F. Guerra, Comm. Math. Phys. \textbf{233}, 1 (2003).

\bibitem{TALE} M. Talagrand, C.R.A.S. \textbf{337}, 111 (2003); Ann. Math. {\bf 163}, 221 (2006).
         
\bibitem{RUELLETREE} D. Ruelle, Comm. Math. Phys. {\bf 48}, 351 (1988).

\bibitem{GG} S. Ghirlanda and F. Guerra, J. Phys.  A: Math.  Gen.  {\bf 31}, 9149 (1998).

\bibitem{AI} M.  Aizenman and P.  Contucci, J. Stat. Phys. {\bf 92},  765 (1998).

\bibitem{SOL} G.  Parisi, Int. J. Mod. Phys. B {\bf 18}, 733 (2004).

\bibitem{VB} L. Viana and A.J. Bray, J. Phys. C {\bf 18}, 3037 (1985).

\bibitem {BETHE} M. M\'ezard, G. Parisi,  Eur. Phys. J. B {\bf 20}, 217 (2001).

\bibitem{ParisiTria} G. Parisi and F. Tria, Eur. Phys. J. B {\bf 30}, 533 (2002).

\bibitem{PaPaRa} A. Pagnani, G. Parisi and M. Rati\'eville, Phys. Rev. E {\bf 68}, 046706 (2003).

\bibitem{TaRiKa} H. Takahashi, F. Ricci-Tersenghi, and Y. Kabashima, Phys. Rev. B {\bf 81}, 174407 (2010).

\bibitem{MaPaRu} E. Marinari, G. Parisi, J. J. Ruiz-Lorenzo, Phys. Rev. {\bf B 58}, 14852 (1998).

\bibitem{Janus5} R. Alvarez Ba–os, A. Cruz, L. A. Fernandez, J. M. Gil-Narvion, A. Gordillo-Guerrero, M. Guidetti, D. I–iguez, A. Maiorano, F. Mantovani, E. Marinari, V. Martin-Mayor, J. Monforte-Garcia, A. Mu–oz-Sudupe, D. Navarro, G. Parisi, S. Perez-Gaviro, F. Ricci-Tersenghi, J. J. Ruiz-Lorenzo, S. F. Schifano, B. Seoane, A. Tarancon, R. Tripiccione, D. Yllanes, {\em Sample-to-sample fluctuations of the overlap distributions in the three-dimensional Edwards-Anderson spin glass}, arXiv:1107.5772  (2011).

\bibitem{Sou} N. Sourlas {\em Universality in Random Systems: the case of the 3-d Random Field Ising model},  arxiv:cond-mat/9810231 (1998).

\bibitem{PaSou} G. Parisi and N. Sourlas, Phys. Rev. Lett. {\bf 89}, 257204 (2002).

\bibitem{PaPiSou} G. Parisi, M. Picco and N. Sourlas, Europhys. Lett. {\bf 66}, 465 (2004).

\bibitem{FraPaVi} S. Franz, G. Parisi and M. Virasoro,  J. Phys. I France {\bf 2}, 1869 (1992).

\bibitem{BiFraMa} A. Billoire, S. Franz and E. Marinari, J. Phys. A {\bf 36} 15 (2003). 

\bibitem{BiCo} A. Billoire and B. Coluzzi, Phys. Rev. E {\bf 67}, 036108 (2003), Phys. Rev. E {\bf 68}, 026131 (2003).

\bibitem{LePaRiRu} L. Leuzzi, G. Parisi, F. Ricci-Tersenghi, and J. J. Ruiz-Lorenzo, Phys. Rev. Lett. {\bf 103}, 267201 (2009).

\end{thebibliography}
\end{document}